\documentclass[preprints,article,accept,moreauthors]{mdpi}
\firstpage{1} 
\pubvolume{1}
\issuenum{1}
\articlenumber{0}
\pubyear{2021}
\copyrightyear{2021}
\externaleditor{Academic Editor: } 
\datereceived{} 
\dateaccepted{} 
\datepublished{} 
\hreflink{https://doi.org/} 
\usepackage{environ}
\NewEnviron{myequation}{%
\begin{equation}
\scalebox{0.88}{$\BODY$}
\end{equation}}

\Title{Strong Coupling Optomechanics Mediated by a Qubit in the Dispersive Regime}

\TitleCitation{Strong Coupling Optomechanics Mediated by a Qubit in the Dispersive Regime}

\Author{Ahmad Shafiei Aporvari $^{1,2}$ 
\orcidA{} and David Vitali $^{1,3,4,}$* 
\orcidB{} }

\AuthorNames{Ahmad Shafiei Aporvari, and David Vitali}

\AuthorCitation{Aporvari, A.S.; Vitali, D.}

\address{%
$^{1}$ \quad School of Science and Technology, Physics Division, University of Camerino, I--62032 Camerino, Italy; ahmad.shafiei@unicam.it\\
$^{2}$ \quad Department of Physics, University of Naples “Federico II”, I-80126 Napoli, Italy \\
$^{3}$ \quad Istituto Nazionale di Fisica Nucleare(INFN) 
, Sezione di Perugia, I--06123 Perugia, Italy \\
$^{4}$ \quad Consiglio Nazionale delle Ricerche--Istituto Nazionale di Ottica (CNR-INO), L.go Enrico Fermi 6, I-50125 Firenze, Italy}

\corres{Correspondence: david.vitali@unicam.it}

\abstract{Cavity optomechanics represents a flexible platform for the implementation of quantum technologies, useful in particular for the realization of quantum interfaces, quantum sensors and quantum information processing. However, the dispersive, radiation–pressure interaction between the mechanical and the electromagnetic modes is typically very weak, harnessing up to now the demonstration of interesting nonlinear dynamics and quantum control at the single photon level. It has  already been shown both theoretically and experimentally that if the interaction is mediated by a Josephson circuit, one can have an effective dynamics corresponding to a huge enhancement of the single-photon optomechanical coupling. Here we analyze in detail this phenomenon in the general case when the cavity mode and the mechanical mode interact via an off-resonant qubit. Using a Schrieffer–Wolff approximation treatment, we determine the regime where this tripartite hybrid system behaves as an effective cavity optomechanical system in the strong coupling regime.}


\begin{document}

\section{Introduction}
Cavity optomechanics~\cite{RMP} has become an established platform for the implementation of quantum information processing in which one can manipulate electromagnetic (e.m.) fields and mechanical/phononic degrees of freedom for the realization of quantum interfaces~\cite{interface1,interface2,interface3}, memories~\cite{memory}, and quantum gates~\cite{gates, bernnett, zhang, asjad}. The optomechanical interaction is typically of parametric form; that is, the cavity frequency is modulated by the motion of the mechanical element, and therefore, it acts dispersively on the e.m. field. However, this interaction is very weak at the level of single quanta because the frequency shift due to a single phonon is typically much smaller than the cavity linewidth and the mechanical frequency. Therefore, one usually operates in the linearized regime where the cavity is intensely driven, and the effective coupling is enhanced by a large intracavity field amplitude~\cite{RMP}. In this latter regime, however, only a limited set of linear operations is possible, harnessing the design of efficient quantum gates within optomechanical platforms. For this reason, there is a growing interest in finding new schemes able to reach the regime where the optomechanical coupling rate $g_{cm}$ is comparable or larger than the cavity decay rate $\kappa$ and the mechanical frequency $\omega_m$. Recently, very promising results have been achieved with new platforms~\cite{arcizet,auffeves}, but it is typically very difficult to achieve simultaneously strong coupling $g_{cm} \geq \omega_m$ and the resolved sideband condition $\kappa < \omega_m$, which is important for enabling coherent control at the quantum level. These latter conditions are instead achievable by adopting a ``hybrid'' approach in which the interaction between the mechanical and the electromagnetic mode is mediated by a qubit simultaneously interacting with both modes~\cite{heikkila,pirkka,Rimberg,steele,tasnimul,chang,northup}. A first experimental proof-of-principle demonstration of coupling enhancement has been achieved using an electromechanical system in a superconducting circuit, where a Cooper pair box~\cite{pirkka} acted as the effective qubit. An alternative hybrid tripartite platform is represented by trapped atoms/ions in an optical cavity~\cite{chang,northup}, in which an internal Raman transition may act as an effective mediator between the cavity field and the atomic motional degree of freedom (see also References~\cite{morigi} for pioneering studies showing the ability to entangle light modes by means of these strongly coupled hybrid tripartite systems). 

The physical mechanism at the basis of the enhancement of the optomechanical coupling, even by many orders of magnitude, is the following. In the dispersive regime in which the qubit is strongly detuned from the cavity (and any driving), the qubit is never excited and remains in its effective ground state. In this case, the AC Stark shift caused by the qubit on the cavity frequency is modulated by the mechanical motion through the qubit-mechanics coupling, resulting in a very strong, effective dispersive optomechanical coupling $g_{cm}^{eff}$, which can reach the strong coupling regime $g_{cm}^{eff} \sim \omega_m$ ($\omega_m$ is the mechanical frequency), provided that the qubit-mechanics coupling rate $g_{am}$ and the qubit-cavity coupling rate $g_{ac}$ are large enough compared to $\omega_m$. 

In this paper, we provide a general treatment of the hybrid tripartite system formed by the cavity mode, the mechanical resonator, and a generic qubit, in the dispersive regime of large detunings, determining the conditions under which the qubit can be adiabatically eliminated, and one can map the tripartite dynamics into that of an effective cavity optomechanical system in the strong coupling regime. The physics of this regime is very different from the one occurring when the qubit-cavity system is quasi-resonant (see for example, Reference~\cite{restrepo} and references therein), and we will exploit the Schrieffer–Wolff method~\cite{SW} in order to arrive at an effective optomechanical model Hamiltonian. We will provide the validity limits of this treatment and the expression of the effective optomechanical coupling rate. We will verify the results by investigating the stationary regime, and in particular, the cavity stationary photon number and the stationary mean phonon number in the low excitation regime, which are suitable for witnessing photon blockade~\cite{Rabl} and other nonlinear optomechanical phenomena in the strong coupling regime~\cite{Nunnenkamp,Law}. 

In Section II, we introduce the tripartite hybrid system and its relevant parameters. In Section III, we describe the Schrieffer–Wolff method through which we derive the effective optomechanical Hamiltonian, while in Section IV, we describe the numerical results showing when the dynamics can be satisfactorily described in terms of a strongly coupled optomechanical system. Section V is for concluding remarks. 

\section{The Hybrid Tripartite System}
\begin{figure}[H]
\includegraphics[width=10.5 cm]{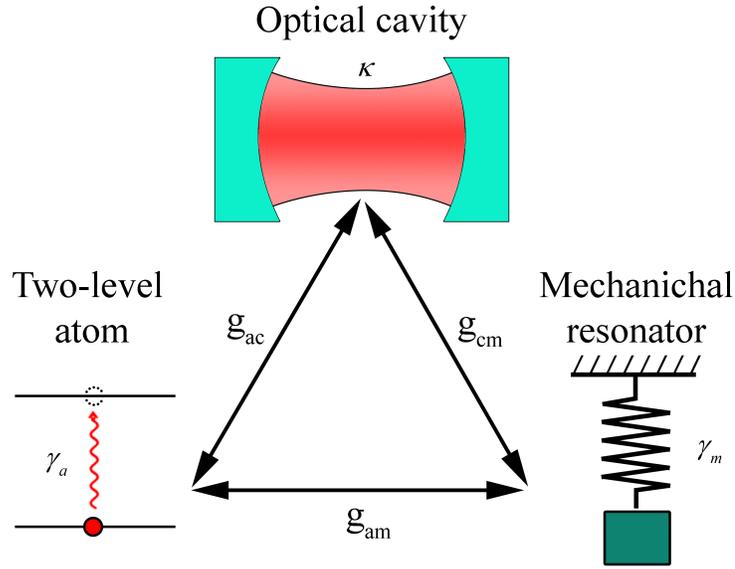}
\caption{{A tripartite hybrid system: A single-mode e.m. cavity, a two-level atom and a mechanical resonator are coupled to each other via the coupling strengths $g_{ac}$, $g_{cm}$, and $g_{am}$. }\label{Fig:System}}
\end{figure}

The tripartite hybrid system we shall study is shown in Figure~\ref{Fig:System}, where a driven single-mode e.m. cavity, a mechanical resonator, and a qubit are mutually coupled. The system Hamiltonian can be quite generally written as ($\hbar =1$)
\begin{equation} \label{Eq:Tot} 
\begin{split}
\hat{H}_t & =  \omega_c \hat{a}^{\dagger} \hat{a} + \frac{1}{2} \omega_a \sigma_z +ig_{ac}\sigma_x (\hat{a}-\hat{a}^{\dagger}) -g_{am} (\hat{\sigma}_{z}+1)(\hat{b}+\hat{b}^\dagger) \\
 & - g_{cm}\hat{a}^\dagger \hat{a}(\hat{b}+\hat{b}^\dagger) +\omega_m \hat{b}^\dagger \hat{b}
 + i F_L (\hat{a} ^\dagger e^{-i\omega_L t} - \hat{a}e^{i\omega_L t}),
\end{split}
\end{equation}
where $\hat{a}$ and $\hat{a}^\dagger$ are the annihilation and creation operators of the cavity mode with frequency $\omega_c$, and $\hat{b}$ and $\hat{b}^\dagger$ those of the mechanical resonator, with frequency $\omega_m$. $\hat{\sigma}_x$, $\hat{\sigma}_y$, and $\hat{\sigma}_z$ are Pauli operators associated with the qubit, whose levels are separated by $\omega_a$. The interaction between the cavity and the qubit with coupling rate $g_{ac}$ is in the full Rabi form, while the qubit--mechanical resonator interaction is of a dispersive nature: the qubit shifts the equilibrium position of the resonator when it is in its (unperturbed) excited state with $\sigma_z = 1$. We also include a direct optomechanical radiation--pressure interaction with coupling rate $g_{cm}$, which is, however, typically much smaller than all the other coupling rates. The last term describes the cavity driving tone, with rate $F_L$ and frequency $\omega_L$; that is, the excitation of the cavity mode through an external classical source, which could be a laser in the optical case or a low noise narrow-band coherent source in the microwave case. The rate $F_L$ is given by $F_L = \sqrt{P_L \eta_{in} \kappa/\hbar \omega_L}$, where $P_L$ is the source power,  $\kappa$ is the cavity decay rate, $0<\eta_{in}\leq 1$ is the mode matching factor between the input driving mode and the cavity mode.

We remark that Equation~(\ref{Eq:Tot}) provides a simplified description of the physical scenario, and in particular, of the system mediating between the electromagnetic cavity and the mechanical resonator, which is here described by a two-level system. In general, one should start from the full electromagnetic interaction between the various subsystems, as for example, in References~\cite{Gudmundsson1, Gudmundsson2}, and consider the whole space of states. However, when the qubit transition frequency $\omega_a$ is clearly separated from all the other transition frequencies of the mediating system, and the driving tone at frequency $\omega_L$ is tuned around $\omega_a$ and is very far from all the other transitions, the present model and the dipole-like interaction assumed in Equation~(\ref{Eq:Tot}) provide a satisfactory description of a wide range of phenomena.

It is convenient to move to the interaction picture with respect to $H_0 = \omega_L (\hat{a}^{\dagger} \hat{a}+\sigma_z/2)$, that is, to move to a frame rotating at the driving tone frequency $\omega_L$, which will represent from now on our frequency reference. In this rotating frame, the counter-rotating terms in the cavity--qubit interaction become $ig_{ac}\left(\sigma_-\hat{a}e^{-2i \omega_L t}-\sigma_+\hat{a}^{\dagger}e^{2i \omega_L t}\right)$, that is, they oscillate at $2\omega_L$, where we have used the usual definitions $\sigma_{\pm}=({\sigma_x}\pm i \sigma_y)/2$. One can make the rotating wave approximation (RWA), i.e., neglect them since they average to zero in the timescales of interest. The resulting total Hamiltonian in this interaction picture therefore becomes 
\begin{equation} \label{Eq:rot} 
\begin{split}
\hat{H}_{hyb} & =  -\Delta \hat{a}^{\dagger} \hat{a} + \frac{1}{2} \Delta_{aL} \sigma_z +ig_{ac}(\sigma_+\hat{a}-\sigma_-\hat{a}^{\dagger}) -g_{am}(\hat{\sigma}_{z}+1)(\hat{b}+\hat{b}^\dagger) \\
 & - g_{cm}\hat{a}^\dagger \hat{a}(\hat{b}+\hat{b}^\dagger) +\omega_m \hat{b}^\dagger \hat{b}
 + i F_L (\hat{a}^\dagger - \hat{a}),
\end{split}
\end{equation}
where $\Delta = \omega_L -\omega_c$ is the cavity detuning and $\Delta_{aL} = \omega_a -\omega_L$ is the atomic detuning from the driving frequency. 

A realistic description of the tripartite system  must also include  decay and noisy processes due to the coupling with the external reservoir. The full description of the dynamics is therefore provided by the following master equation for the density matrix $\hat{\varrho}_t$ of the whole tripartite system
\begin{equation} \label{Eq:master-Hyb} 
\begin{split}
\frac{d\hat{\varrho}_t}{dt} & =-i\left[\hat{H}_{hyb},\hat{\varrho}_t\right]  +\kappa {\cal D}[\hat{a}] \hat{\varrho}_t +\gamma_a {\cal D}[\hat{\sigma}_-] \hat{\varrho}_t \\
 &  +\gamma_m (n_{th}+1){\cal D}[\hat{b}] \hat{\varrho}_t +\gamma_m n_{th}{\cal D}[\hat{b}^{\dagger}] \hat{\varrho}_t,
\end{split}
\end{equation}
where ${\cal D}[\hat{c}] \hat{\varrho}_t=\hat{c} \hat{\varrho}_t\hat{c}^{\dagger}- (\hat{c}^{\dagger} \hat{c}\hat{\varrho}_t+\hat{\varrho}_t \hat{c}^{\dagger} \hat{c})/2$ is the standard dissipator in Lindblad form, $\kappa$ is the cavity decay rate, $\gamma_a$ the qubit decay rate from the excited to the ground state, $\gamma_m$ is the mechanical damping rate, and $n_{th}=(e^{\hbar \omega_m/k_B T}-1)^{-1}$ is the mean thermal excitation number of the reservoir of the mechanical mode. We consider thermal equilibrium at temperature $T$, and thermal excitations (and the corresponding Lindblad terms) are negligible for the qubit and cavity subsystems because $\hbar {\omega_c}/k_B T \sim \hbar {\omega_a}/k_B T  \gg 1$.

In the next section, we will show that when the qubit is far off resonance from the cavity and its driving, it is able to mediate an effective dispersive interaction between the cavity and the mechanical resonator, reproducing therefore an effective optomechanical system in the strong coupling regime.

\section{The Schrieffer--Wolff Approximation and Effective Optomechanical System}

We first rewrite Equation~(\ref{Eq:rot}) by grouping together the terms involving the qubit operators, 
\begin{equation} \label{Eq:rot2} 
\hat{H}_{hyb}  =  \hat{H}_{qubit}-\Delta \hat{a}^{\dagger} \hat{a}  + i F_L (\hat{a}^\dagger - \hat{a})  +\omega_m \hat{b}^\dagger \hat{b}-(g_{am}+g_{cm}\hat{a}^\dagger \hat{a})(\hat{b}+\hat{b}^\dagger).
\end{equation}
$\hat{H}_{qubit}$ can be written as that of a magnetic dipole in an effective magnetic field,
\begin{equation} \label{Eq:qubit} 
\begin{split}
\hat{H}_{qubit} & =  \frac{1}{2}\left(\hat{B}_x \hat{\sigma}_y+\hat{B}_y \hat{\sigma}_y+\hat{B}_z \hat{\sigma}_z\right),
\end{split}
\end{equation}
where $\hat{B}_x=-g_{ac}\hat{p}_c$, $\hat{B}_y=-g_{ac}\hat{x}_c$, $\hat{B}_z=\Delta_{aL}-2g_{am}\hat{x}_m$, with $p_c =-i(a-a^{\dagger})$, $x_c =a+a^{\dagger}$, and $x_m =b+b^{\dagger}$.

We now make the important assumption that $\Delta_{aL}$ is larger than the coupling rates of the qubit, $g_{ac}$ and $g_{am}$, that is, the qubit is far off-resonance from the cavity and the mechanical resonator, and it is not excited by the cavity driving. In this dispersive limit, the qubit does not exchange energy with the other subsystems, and it remains in its effective ground state. We are in the condition to apply the Schrieffer--Wolff method because we have a lower energy subspace, corresponding to the effective qubit ground state, which is well separated from the high energy subspace. In this lower energy subspace, the effective qubit Hamiltonian is
\begin{equation} \label{Eq:qubiteff} 
\hat{H}_{qubit}^{eff} =  -\frac{1}{2}\sqrt{\hat{B}_x^2 +\hat{B}_y^2+\hat{B}_z ^2}=-\frac{1}{2}\sqrt{4 g_{ac}^2\left(\hat{a}^{\dagger} \hat{a} +\frac{1}{2}\right)+\left(\Delta_{aL}-2 g_{am} \hat{x}_m\right)^2},
\end{equation}
which is an effective operator acting on the Hilbert space of the optomechanical system. It is now consistent to expand this effective square-root operator as a power series in the small parameters $g_j/\Delta_{aL}$ $(g_j = g_{ac},g_{am})$, and we stop at the third order, using $\sqrt{1+\eta}\simeq 1+\eta/2-\eta^2/8+\eta^3/16$. One gets
\begin{equation} \label{Eq:qubiteff2} 
\hat{H}_{qubit}^{eff} =  -\frac{\Delta_{aL}}{2} +g_{am}(\hat{b}+\hat{b}^{\dagger})- \frac{g_{ac}^2}{\Delta_{aL}}\left(\hat{a}^{\dagger} \hat{a} +\frac{1}{2}\right)-\frac{2g^2_{ac}g_{am}}{\Delta^2_{aL}}\left(\hat{a}^{\dagger} \hat{a} +\frac{1}{2}\right)(\hat{b}+\hat{b}^\dagger),
\end{equation}
which, inserted into Equation~(\ref{Eq:rot2}) and neglecting constant energy terms, yields the following effective, low-energy, optomechanical Hamiltonian valid in the considered dispersive regime for the qubit,
\begin{myequation} \label{Eq:eff} 
\hat{H}_{om}^{eff} =  -(\Delta + \frac{g^2_{ac}}{\Delta_{aL}}) \hat{a}^{\dagger} \hat{a}  -(\frac{g^{eff}_{cm}- g_{cm}}{2})(\hat{b}+\hat{b}^\dagger)
  - g^{eff}_{cm}\hat{a}^\dagger \hat{a}(\hat{b}+\hat{b}^\dagger) +\omega_m \hat{b}^\dagger \hat{b}
 + i F_L (\hat{a}^\dagger - \hat{a}),
\end{myequation}
where 
\begin{equation} \label{Eq:geff}
g^{eff}_{cm} = g_{cm}+ \frac{2g^2_{ac}g_{am}}{\Delta^2_{aL}}
\end{equation}
is the effective optomechanical radiation--pressure-like interaction rate, with the additional term at third order in $g_j/\Delta_{aL}$ representing the effective indirect interaction mediated by the qubit through the AC Stark shift. This latter term can be significantly larger than the direct optomechanical coupling $g_{cm}$, and under the condition $\Delta_{aL} \gg g_{ac}, g_{am} \gg \omega_m$, one expects to achieve the strong coupling regime $g^{eff}_{cm} \sim \omega_m$. 
In order to verify this fact, we have to therefore  compare the dynamics of the full tripartite system associated with \mbox{Equation~(\ref{Eq:master-Hyb})} to that of the effective optomechanical Hamiltonian described by the following master equation for the cavity-mechanics density operator $\varrho$ 
\begin{equation} \label{Eq:master-eff} 
\frac{d\hat{\varrho}}{dt} =-i\left[\hat{H}_{om}^{eff},\hat{\varrho}\right]  +\kappa {\cal D}[\hat{a}] \hat{\varrho}  +\gamma_m (n_{th}+1){\cal D}[\hat{b}] \hat{\varrho} +\gamma_m n_{th}{\cal D}[\hat{b}^{\dagger}] \hat{\varrho}.
\end{equation}

\section{Results for the Stationary State of the Optomechanical System}

\begin{figure}[H]
\includegraphics[width=13.5 cm]{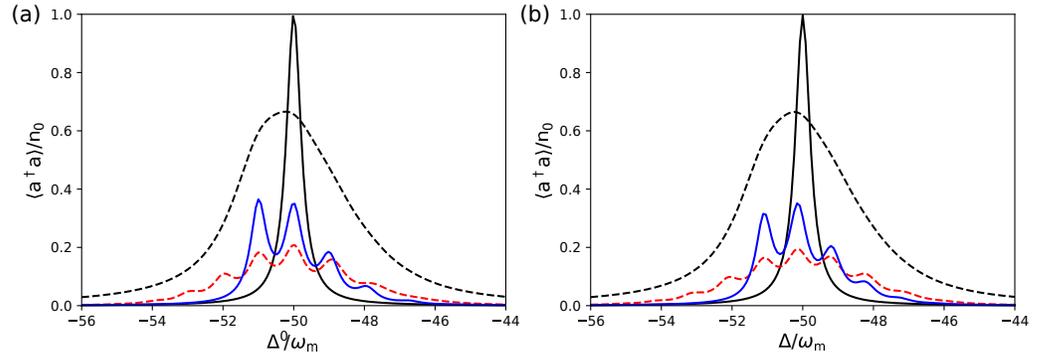}
\caption{Stationary cavity photon number $ \langle \hat{a}^\dagger \hat{a}\rangle$ versus the cavity detuning for the effective optomechanical model of Equation~(\protect\ref{Eq:master-eff}) (\textbf{a}), and for the hybrid tripartite system with master equation Equation~(\protect\ref{Eq:master-Hyb}) (\textbf{b}). We have chosen the following set of parameters: $\omega_a/\omega_m = 1.5 \times 10^4$,  $\omega_L/\omega_m = 10^4 $, $g_{am}/\omega_m=50$, $g_{ac}/\omega_m=500 $, $g_{cm}/\omega_m = 10^{-3}$, so that the system is in the strong optomechanical coupling regime $g^{eff}_{cm}= \omega_m$. In (\textbf{a}), we use the effective detuning $\Delta ' =\Delta + g^{eff}_{cm}(g^{eff}_{cm}-g_{cm})/\omega_m$, which takes into account the cavity frequency shift associated with the mechanical displacement in Equation~(\protect\ref{Eq:eff}). The other parameters are $\omega_m/\kappa= 2$ and $n_{th}=0$ for the blue solid line, $\omega_m/\kappa= 2$ and $n_{th}=1$ for the red dashed line, $\omega_m/\kappa= 0.5$ and $n_{th}=0$ for the black dashed line. The black solid line refers to the uncoupled cavity, i.e., $g_{j}= 0$ (where $j=am, ac$ and $cm$). All curves are normalized with respect to the value of the peak of this latter curve, $n_0=4F^2_L/\kappa^2$. In this parameter regime, the two models provide almost indistinguishable predictions. We have also taken $\gamma_m/\omega_m=\gamma_a/\omega_m =1/20$, and $F_L = 10^{-2} \sqrt{\kappa}$.\label{Fig:n_set1}}
\end{figure}

\begin{figure}[H]
\widefigure
\includegraphics[width=15 cm]{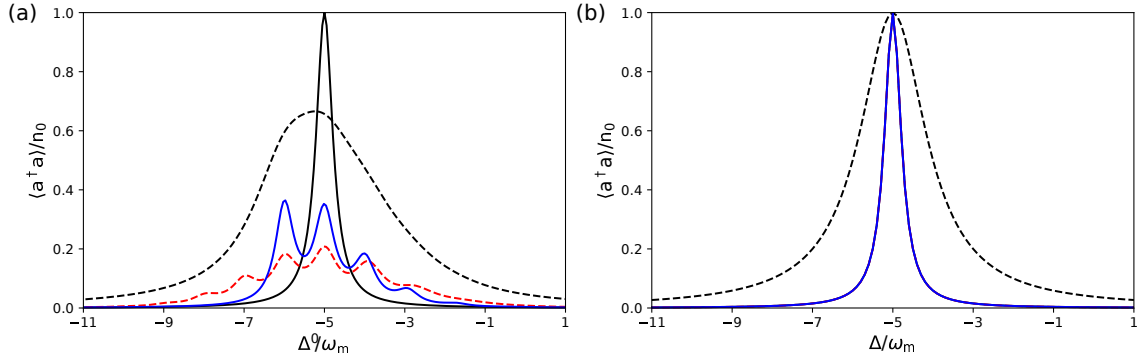}
\caption{Stationary cavity photon number $ \langle \hat{a}^\dagger \hat{a}\rangle$ versus the cavity detuning for the effective optomechanical model of Equation~(\protect\ref{Eq:master-eff}) (\textbf{a}), and for the hybrid tripartite system with the master equation, Equation~(\protect\ref{Eq:master-Hyb}) (\textbf{b}), for the following set of parameters, different from that of Figure~\ref{Fig:n_set1}; $\omega_a/\omega_m = 1.5 \times 10^3$,  $\omega_L/\omega_m = 10^3 $,  $g_{am}/\omega_m=g_{ac}/\omega_m = 50$, $g_{cm}/\omega_m = 10^{-3}$. Furthermore, with these different values, the strong optomechanical coupling regime condition, $g^{eff}_{cm}= \omega_m$, holds. The other parameters and styles of the curve are the same as in Figure~\ref{Fig:n_set1}. In this parameter regime, the assumptions of the Schrieffer--Wolff method are only approximately valid, and the two models provide different predictions. \label{Fig:n_set2}}
\end{figure}

Here we focus on the stationary state of the system achieved at long times. We first consider the stationary cavity photon number $ \langle \hat{a}^\dagger \hat{a}\rangle$, and the results for the two dynamics are compared in Figures~\ref{Fig:n_set1} and \ref{Fig:n_set2}, where we study the behavior of $ \langle \hat{a}^\dagger \hat{a}\rangle$ as a function of the cavity detuning.

In Figure~\ref{Fig:n_set1}, we consider a set of parameters satisfying the dispersive regime described in the previous section, $\Delta_{aL} \gg g_{ac}, g_{am} \gg \omega_m$, where the tripartite hybrid system  reproduces a strongly coupled optomechanical system well, that is, $\omega_a/\omega_m = 1.5 \times 10^4$,  $\omega_L/\omega_m = 10^4 $, $g_{am}/\omega_m=50$, $g_{ac}/\omega_m=500 $, $g_{cm}/\omega_m = 10^{-3}$. In this case, in fact, $g^{eff}_{cm}= \omega_m$, and, as predicted, the two master equations, Equations~(\ref{Eq:master-Hyb}) and~(\ref{Eq:master-eff}), yield almost indistinguishable predictions. Moreover, the typical signatures of strong optomechanical coupling manifest themselves because we see that, in the weak excitation limit of small driving rate $F_L$, and in the resolved sideband regime $\kappa < \omega_m$ (the blue and red curves of Figure~\ref{Fig:n_set1}), the resonance peaks corresponding to the absorption of single mechanical quanta are clearly visible \cite{Nunnenkamp}. At a finite thermal phonon number $n_{th}$ (see the red curves and the caption of Figure~\ref{Fig:n_set1}), additional peaks appear even though for increasing $n_{th}$ they tend to blur into a broad thermal background \cite{Nunnenkamp}. The various resonances overlap and vanish as soon as we move to the unresolved sideband regime $\kappa > \omega_m$ (dashed black lines in Figure~\ref{Fig:n_set1}), and we get a broad peak, even larger than the standard Lorenztian response of the cavity, which tends to be reproduced at strong driving and not too strong coupling. We notice that thanks to the mediating action of the off-resonant qubit, the optomechanical coupling has been increased by three orders of magnitude. We also notice that in \mbox{Figure~\ref{Fig:n_set1}a}, we use the effective detuning $\Delta ' =\Delta + g^{eff}_{cm}(g^{eff}_{cm}-g_{cm})/\omega_m$ rather than $\Delta$, in order to take into account the cavity frequency shift associated with the mechanical displacement term, $ -(\frac{g^{eff}_{cm}- g_{cm}}{2})(\hat{b}+\hat{b}^\dagger)$ in Equation~(\ref{Eq:eff}). All the other parameters are given in the figure caption.

\end{paracol}
\nointerlineskip
\begin{paracol}{2}
\switchcolumn

In Figure~\ref{Fig:n_set2}, we consider a slightly different set of parameters, $\omega_a/\omega_m = 1.5 \times 10^3$,  $\omega_L/\omega_m = 10^3 $,  $g_{am}/\omega_m=g_{ac}/\omega_m = 50$, $g_{cm}/\omega_m = 10^{-3}$, which again satisfies the strong optomechanical coupling regime condition, $g^{eff}_{cm}= \omega_m$, while all the other parameters are kept unchanged. This figure shows that the equivalence between the two models of Equations~(\ref{Eq:master-Hyb}) and~(\ref{Eq:master-eff}) is not easy to achieve, and it is valid in a quite limited parameter region. In fact, even though  we apparently still satisfy the conditions for the Schrieffer--Wolff method because $g_j/\Delta_{aL} =0.1$, we see that we have very different predictions for the cavity photon number versus detuning. The prediction of the effective optomechanical model of Equation~(\ref{Eq:master-eff}) is almost unchanged, while that of the full tripartite system is now very different, it shows no additional resonance peaks, and it does not differ significantly from the standard Lorenztian form. A closer inspection of the chosen parameters explains why in this latter case the two models significantly differ. In fact, the effective optomechanical Hamiltonian of Equation~(\ref{Eq:eff}) is valid up to \emph{first} order in $g_{am}/\Delta_{aL}$, and up to \emph{second} order in $g_{ac}/\Delta_{aL}$. Therefore, one needs to consider \emph{smaller} values of $g_{am}$ compared to $g_{ac}$ due to the lower accuracy in the expansion parameter $g_{am}/\Delta_{aL}$. This is verified for the set of parameters of  Figure~\ref{Fig:n_set1}, where $g_{am}/\Delta_{aL}=0.01$ and  $g_{ac}/\Delta_{aL}=0.1$, and it is not satisfied for the choice of parameters of Figure~\ref{Fig:n_set2}, for which $g_{am}/\Delta_{aL}=g_{ac}/\Delta_{aL}=0.1$. This higher value of the qubit--mechanics coupling alone is responsible for a very different behavior of the stationary state of the system.  

These findings are confirmed by the behavior of another steady-state quantity, the mean phonon number of the mechanical resonator $ \langle \hat{b}^\dagger \hat{b}\rangle$, which we study again as a function of detuning, and it is shown in Figures~\ref{Fig:b_set1} and \ref{Fig:b_set2}. 

\begin{figure}[H]
\includegraphics[width=13.5 cm]{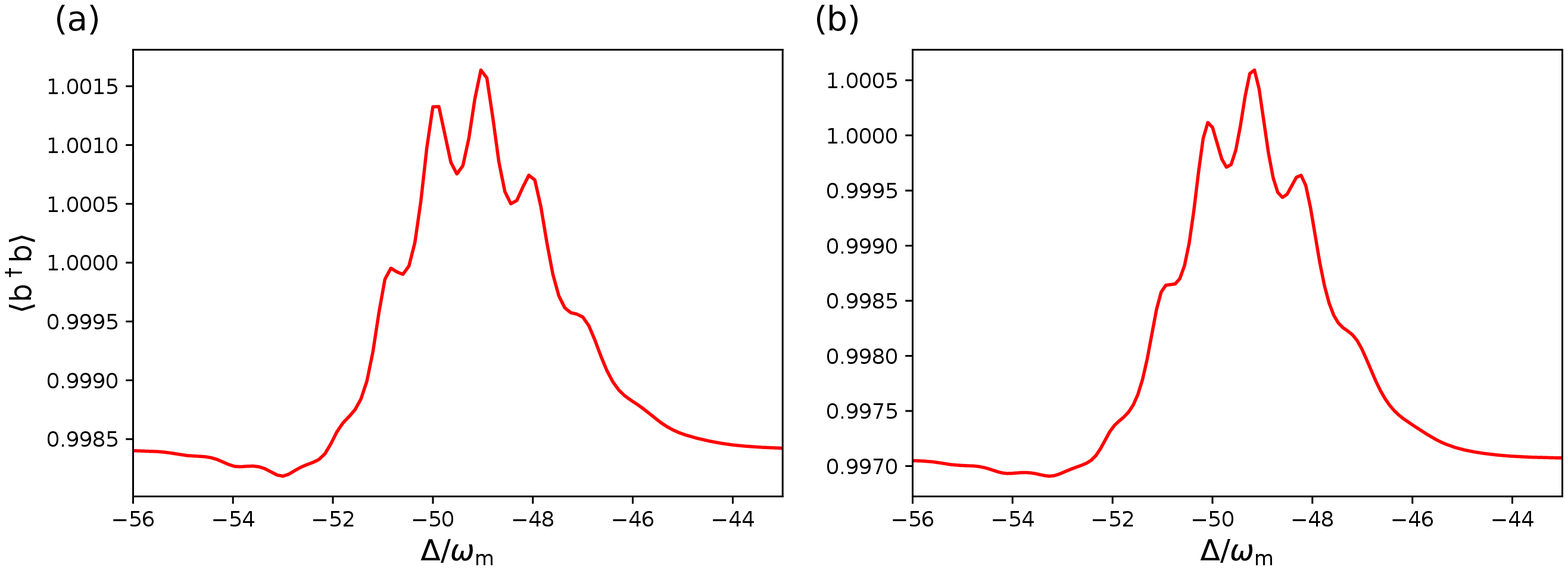}
\caption{Stationary cavity phonon number $ \langle \hat{b}^\dagger \hat{b}\rangle$ versus the cavity detuning for the effective optomechanical model of Equation~(\protect\ref{Eq:master-eff}) (\textbf{a}), and for the hybrid tripartite system with the master equation, Equation~(\protect\ref{Eq:master-Hyb}) (\textbf{b}). For this curve, we used the same parameters as in the red dashed line in Figure \ref{Fig:n_set1}:  $\omega_m/\kappa= 2$ and $n_{th}=1$.  We also choose the set of parameters as in Figure~\protect\ref{Fig:n_set1}: $\omega_a/\omega_m = 1.5 \times 10^4$,  $\omega_L/\omega_m = 10^4 $, $g_{am}/\omega_m=50$, $g_{ac}/\omega_m=500 $, $g_{cm}/\omega_m = 10^{-3}$,  which is the parameter regime where the Schrieffer--Wolf approximation is valid. The other parameters are as in Figure \ref{Fig:n_set1}.\label{Fig:b_set1}}
\end{figure}
\vspace{-6pt}
\begin{figure}[H]
\includegraphics[width=13.5 cm]{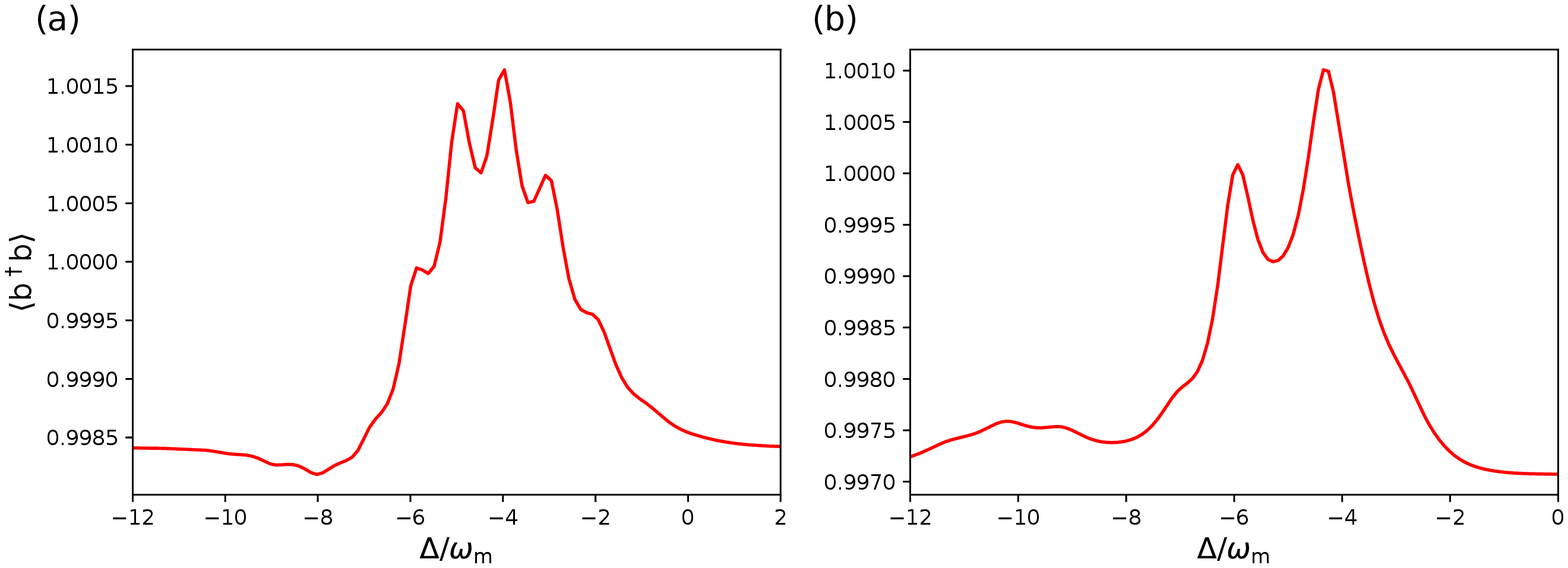}
\caption{Stationary cavity phonon number $ \langle \hat{b}^\dagger \hat{b}\rangle$ versus the cavity detuning for the effective optomechanical model of Equation~(\protect\ref{Eq:master-eff}) (\textbf{a}), and for the hybrid tripartite system with master equation Equation~(\protect\ref{Eq:master-Hyb}) (\textbf{b}). The parameters are the same as in Figure~\protect\ref{Fig:n_set2}: $\omega_a/\omega_m = 1.5 \times 10^3$,  $\omega_L/\omega_m = 10^3 $,  $g_{am}/\omega_m=g_{ac}/\omega_m = 50$, $g_{cm}/\omega_m = 10^{-3}$ where the predictions of the two models are different. The other parameters are as in Figure \ref{Fig:b_set1}. \label{Fig:b_set2}}
\end{figure}

In Figure~\ref{Fig:b_set1}, we consider the same parameters of Figure~\ref{Fig:n_set1} and focus on the case corresponding to the red dashed curve ($\omega_m/\kappa=2$ and $n_{th} =1$). At this low temperature, the stationary phonon number $ \langle \hat{b}^\dagger \hat{b}\rangle$ is only weakly modified by the weak cavity driving, but the resonances corresponding to the various phonon transitions are clearly visible, and the two models provide very similar predictions. On the contrary, in Figure~\ref{Fig:b_set2}, we consider the same parameters of Figure~\ref{Fig:n_set2}, and also for the stationary mechanical excitation, in this different parameter regime in which $g_{am}/\omega_m$ is not small enough, the full hybrid tripartite system of Equation~(\ref{Eq:master-Hyb}) and the effective optomechanical model of Equation~(\ref{Eq:master-eff}) provide clearly distinct behavior.

\section{Discussion}

In this paper, we have shown under which conditions a qubit can effectively mediate the interaction between a mechanical resonator and an electromagnetic cavity mode, enabling to reach the optomechanical strong-coupling limit, which is otherwise very difficult to achieve in conventional optomechanical systems~\cite{RMP,arcizet}. We have reconsidered in general this idea that was put forward within the context of superconducting electromechanical systems~\cite{heikkila,pirkka,Rimberg,steele,tasnimul} and trapped atoms ~\cite{chang,northup}. We have seen that in the dispersive regime when the qubit is far off resonance so that the condition $\Delta_{aL} \gg g_{ac}, g_{am} \gg \omega_m$ is satisfied, there is a parameter regime where the qubit-cavity-mechanical resonator system behaves very similarly to an optomechanical system in the strong coupling regime, where the effective optomechanical coupling rate is comparable to the frequency. We have verified this fact by looking at the stationary properties of the cavity and of the mechanical resonator. The parameter regime in which the dynamics driven by Equations~(\ref{Eq:master-Hyb}) and~(\ref{Eq:master-eff}) is equivalent is, however, quite limited because this is valid at first order in $g_{am}/\Delta_{aL}$ and at second order in $g_{ac}/\Delta_{aL}$.

Despite the limited validity range of our treatment, the present optomechanical coupling enhancement could be designed and tested in the case of superconducting circuits coupled to driven microwave cavities. A proof-of-principle demonstration in these setups has already been  given in Reference~\cite{pirkka}. A more effective and clear demonstration could be given using circuits with a very large charging energy so that the qubit transition frequency $\omega_a$ is clearly separated from all the other transition frequencies of the circuit. Then, for example, the conditions of Figure~\ref{Fig:n_set1} could be realistically implemented taking achievable values such as $\omega_m = 1$ MHz, $\omega_a = 15$ GHz, $\omega_L = 10$ GHz, $g_{am} = 50$ MHz, \mbox{$g_{ac} = 500$ MHz}. Values of $\kappa$ equal to 0.5 or 2 MHz are easy to achieve for a microwave cavity, and one should operate in a dilution fridge environment. Furthermore, weak driving with small intracavity photon number $n_0$ is achieved by using attenuators.

The present study can be extended in various directions. One can compare the two models in more detail by also looking at dynamical quantities such as spectra and optomechanical correlations, and also focus on the regime where other higher-order nonlinear phenomena, such as the cross--Kerr interaction, play a role due to the presence of the mediating qubit~\cite{pirkka,tasnimul}. Another interesting option for quantum information applications is to consider the case when the qubit mediates the interaction between two (or more) mechanical and electromagnetic modes in order to exploit the strong coupling (and also the resolved sideband) regime for the realization of quantum gates between photonic and phononic qubits~\cite{bernnett}. 
\vspace{6pt}

\authorcontributions{Conceptualization, David Vitali; Data curation, Ahmad Shafiei Aporvari; Methodology, David Vitali; Software, Ahmad Shafiei Aporvari; Supervision, David Vitali; Writing – original draft, Ahmad Shafiei Aporvari, David Vitali. All authors have read and agreed to the published version of the manuscript.}

\funding{This research received no external funding.}

\dataavailability{ The data that support the findings of this study are available on request from the corresponding author. Data and simulation code are available at \url{https://github.com/ahmad-shafiei/Triparti_System.git}. }


\conflictsofinterest{The authors declare no conflict of interest.} 


\end{paracol}
\reftitle{References}




\end{document}